\documentclass[prd,aps,twocolumn]{revtex4-1}

\usepackage{graphicx} 
\usepackage{amsmath}
\usepackage{amssymb}
\usepackage{dcolumn}% Align table columns on decimal point
\usepackage{bm}% bold math
\usepackage{color}

\newcommand{\xpom}{x_{I\!\!P}}
\newcommand{\dint}{{\rm d}}

\begin{document}

\title{Exclusive diffractive processes in electron-ion collisions}

\author{Tobias Toll}
\email[]{ttoll@bnl.gov}
\affiliation{Brookhaven National Laboratory, Upton, NY}
\author{Thomas Ullrich}
\email[]{thomas.ullrich@bnl.gov}
\affiliation{Brookhaven National Laboratory, Upton, NY}

\date{\today}

\begin{abstract}
    We present a new technique to calculate the cross-section for
    diffractive vector meson production and DVCS in electron-ion
    collisions based on the dipole model.  The measurement of these
    processes can provide valuable information on non-linear QCD
    phenomena, such as gluon saturation, and is the the only known way
    to gain insight into the spatial distribution of gluons in nuclei.
    We present predictions of differential cross-section distribution
    $d\sigma/dQ^2$ and $d\sigma/dt$ for $J/\psi$ and $\phi$ meson
    production for diffractive processes of heavy nuclei and
    demonstrate the feasibility of extracting the gluon source
    distribution of heavy nuclei, $F(b)$, from coherent diffraction. We
    briefly introduce a new event generator based on our method that
    can be used for studying exclusive diffractive processes at a
    future electron-ion collider.
\end{abstract}


\maketitle %\maketitle must follow title, authors, abstract and \pacs

\section{Introduction}
%
% HERA legacy, saturation
%
The HERA accelerator at DESY, Germany, with collision energies of
$\sqrt{s} = 320$ GeV was the hitherto highest energy lepton-hadron
collider.  One of the great achievements of HERA was the determination
of the partonic structure of the proton \cite{Aaron:2009aa}.
% the discovery of the importance of gluons in the structure of the
% proton and the detailed study of the running of the strong coupling.
A lepton-hadron collision is mediated by a virtual photon, which
interacts with a valence- or sea-quark within the hadron at a
resolution $Q^2$.  When probed at higher energies, gluons fluctuating
into gluon- or quark-pairs can be resolved at smaller time scales,
such that more partons share the hadron's longitudinal momentum at
higher energies.  At small momentum fractions $x \lesssim 10^{-2}$ of
the participating partons, measurements at HERA showed that the
content of the proton is dominated by gluons, and that the gluon
number density
% s' combined momentum fraction
at smaller $x$ seems to rise uncontrollably.  When extrapolating
current measurements to small $x$-values, 
the gluonic part of the cross-section becomes larger than
the total proton cross-section.
%the combined longitudinal
%momentum carried by all gluons in the proton exceeds that of the
%proton itself.
% This is a consequence of the linear QCD evolution, which is able to
% describe the HERA data perfectly.
This violation of the unitarity bound can only be avoided by
introducing saturation effects that tame the explosive growth of the
gluon density.  While many saturation models describing these
non-linear effects were developed \cite{Iancu:2003xm, Weigert:2005us},
there exists no {\it direct} measurement that would allow to verify
these models and ultimately prove the existence of gluon saturation.
Although more and more tantalizing hints of the onset of gluon
saturation coming from proton-ion collisions at RHIC have become available
\cite{Arsene:2004ux, Adams:2006uz, Braidot:2010ig, Adare:2011sc,
    Albacete:2010pg, Lappi:2012nh}, alternative explanations can
currently not be ruled out \cite{Strikman:2010bg, Kang:2011bp,
    Kang:2012kc}.  The direct study of these non-linear saturation
effects would require lepton-hadron collisions at energies far
exceeding those at HERA.  Electron-ion collisions offer an alternative
way to study high gluon-density phenomena at an order of magnitude
lower center-of-mass energies.
% if one instead of protons considers heavy ions accelerated to high
% energies.
At high enough energies the small $x$ gluons in the
heavy ion have a wave length in the longitudinal direction that
encompass the entire width of the nucleus. 
%Thus, at a given point
%in the transverse plane, the number-density of gluons becomes
%proportional to the nucleus's longitudinal thickness in this point. 
A probe will thus coherently interact with the bulk of low-$x$ gluons.
For a heavy ion, the thickness is approximately constant away from the
edges and is proportional to $A^{1/3}$, where $A$ is its atomic
number.  This approximate dependence is supported by detailed studies
\cite{Kowalski:2007rw, Kowalski:2003hm}.  Therefore, probing a heavy
ion with $A\approx 200$ is similar to probing a proton at 6 times
higher energy, making the nucleus an efficient amplifier of the physics
of high gluon densities.

There are two proposed future collider projects that aim to directly
measure the saturated gluon regime for the first time: the Large
Hadron-electron Collider (LHeC) at CERN using the existing LHC machine
\cite{AbelleiraFernandez:2012cc} and the Electron-Ion Collider (EIC)
in the USA \cite{Deshpande:2012bu}, using either the existing RHIC accelerator
complex at BNL (eRHIC), or the existing electron beams at JLab (MEIC).

%
% HERA legacy, diffraction
%
At HERA, an unexpected discovery was that approximately 10\% of the
$ep$ cross-section is from diffractive final states
\cite{Abramowicz:1998ii} and that this fraction is fairly independent
of $W$ and $Q^2$.  What characterizes these events experimentally is
the presence of a rapidity gap, a region in the angular coverage which
exhibits {\it no} hadronic activity.  Diffractive interactions result
when the electron probe in Deeply Inelastic Scattering (DIS) 
interacts with a color neutral vacuum
excitation. This vacuum excitation, which in perturbative QCD may be
visualized as a colorless combination of two or more gluons, is often
called the Pomeron.  The hard diffractive cross-section is
proportional to the gluon-density {\it squared}, making it the most
sensitive probe of gluon density known. Thus, diffraction and
saturation are closely related phenomena.

%
% RHIC legacy, initial conditions
%
Measurements of diffraction in an electron-ion collider also have
substantial potential to shed light on other unanswered questions in
heavy ion collision \cite{Deshpande:2012bu}.  Measurements over the last decade
in heavy ion collision experiments at RHIC indicate the formation of a
strongly coupled plasma of quarks and gluons (sQGP).  This sQGP
appears to behave like a ``near-perfect liquid" with a ratio of the
shear viscosity to entropy density ($\eta/s$) approaching $1/4\pi$
\cite{Adcox:2004mh, Adams:2005dq, Back:2004je, Arsene:2004fa}.  Recent
experiments at the LHC with substantially higher energies and thus a
hotter and longer lived plasma phase confirm this picture
\cite{Tserruya:2011dy}.  Despite the significant insight that the sQGP
is a strongly correlated nearly perfect liquid, little is understood
about how the system is created.
% and what its properties ($\eta/s(T)$,
%transport coefficients, etc) are.  
The largest uncertainty in our
understanding of the evolution of a heavy-ion collision comes from our
limited knowledge of the initial condition, i.e. momentum and spatial
distributions of gluons in the nuclei.  Also of importance is how the
spatial distribution fluctuates around its mean, since it affects the
behavior of collective effects such as flow and their higher moments.
For example, different assumptions about the nuclear initial
distributions give differences up to factors of two for the obtained
$\eta/s$ value \cite{Qiu:2011iv, Esumi:2011nd}.  Measurements of the
initial gluon distribution with existing machines are only possible
indirectly and with large uncertainties.  The study of gluon
distributions using diffractive events in electron-ion collisions
would allow one to directly measure the initial condition of the
colliding ions, providing both its momentum and spatial distributions
as well as the underlying fluctuations (``lumpiness''). In fact, exclusive
diffractive $e$A events are the only way to study the initial spatial
distributions and shed light on these fundamental questions.

In a diffractive $e$A event, the electron collides with the ion
producing one or more extra particles but leaving the nucleus intact.
The interaction with the nucleus is either elastic or inelastic, and
in the latter case the nucleus subsequently radiates a photon or breaks
up into color neutral fragments. When it stays intact, the event is
called coherent and when it breaks, the event is called incoherent.
The spectrum of the cross-section with respect to the hadronic
momentum transfer $t$ is related to the transverse spatial
distribution of the gluons in the ion through a Fourier transform.
Also, according to the Good-Walker picture \cite{Good:1960ba}, the
incoherent cross-section is a direct measure of the lumpiness of the
gluons in the ion.  In order to access $t$ in these events, the
complete final state has to be measured. This is experimentally only
possible in events such as vector-meson production or Deeply Virtual
Compton Scattering (DVCS).

%
% Dipole models
%
At present, the most common approach to calculate diffractive
cross-sections at small $x$ is in the dipole model, where the
exchanged virtual photon splits up into a quark anti-quark pair, which
forms a color dipole.  The dipole subsequently interacts with the
target in the target's rest frame.  The dipole model became an
important tool for DIS when Golec-Biernat and W\"usthoff (GBW)
\cite{GolecBiernat:1998js, GolecBiernat:1999qd} observed that a simple
ansatz for the dipole model integrated over the impact parameter was
able to simultaneously describe the total inclusive and diffractive
cross-sections. The GBW model also naturally contains saturation in
the small $x$ regime.  A shortcoming of the GBW model is that it
cannot describe the high $Q^2$ scaling violation in the inclusive
cross-sections measured at HERA, something perfectly described by the
collinear DGLAP formalism, which in turn cannot describe the high
fraction of diffractive events.  This sparked Bartels, Golec-Biernat
and Kowalski (BGBK) to include an explicit DGLAP gluon distribution
into the dipole formalism \cite{Bartels:2002cj}, taken at a scale
directly linked to the dipole size.  The BGBK model replicates the GBW
model where it is applicable and also manages do describe the $Q^2$
dependence of the cross-sections. However, this approach still
integrates out the impact parameter dependence of the interaction,
without which the $t$-dependence of the cross-section is unknown.  The
impact parameter dependence was introduced in the dipole model by
Kowalski and Teaney \cite{Kowalski:2003hm} and then modified to also
include exclusive processes by Kowalski, Teaney and Motyka
\cite{Kowalski:2006hc}. This dipole model goes by the name bSat (or
sometimes IPSat), and is the main focus of this paper.

The bSat model has been studied in detail in the case of electron
proton collisions at HERA.  There are a few theoretical attempts to
expand the bSat model to also describe exclusive $e$A collisions
 (see e.g.~\cite{Kowalski:2003hm, Caldwell:2010zza, Lappi:2010dd}).  
Without exception, these models
fail to describe the disappearance of the incoherent cross-section
as $t\rightarrow 0$.
%contain rather large approximations when
%describing the incoherent cross-section
%These description are accurate for $|t|\gg 0$. 
%As $|t|\rightarrow 0$ the incoherent part of the cross-section is expected
%to approach naught, something previous descriptions have not encompassed.  
Also, they turn out to be poorly suited for
implementation in a Monte Carlo event generator.

In this paper we present the first calculations of not only the
coherent but also the incoherent cross-sections in electron-ion
collision without making approximations larger than those already
inherently present in the bSat model, for all $t$.  
We have implemented the
calculation described in this paper in a Monte Carlo event generator
(Sar{\it t}re).

The paper is organized as followed: In section \ref{bSat} we will show
our derivation of the dipole model in $e$A, taking as a starting point
the case of $ep$. In section \ref{results} we will present the
resulting cross-sections, both as comparisons with HERA data and as
predictions for EIC and RHIC.

%%%%%%%%%%%%%%%%%%%%%%%%%%%%%%%%%%%%%%%%%%%%%%%%%%%%%%%%%%%%%%%%%%%%%%%%
\section{The bSat dipole model}
\label{bSat}
%%%%%%%%%%%%%%%%%%%%%%%%%%%%%%%%%%%%%%%%%%%%%%%%%%%%%%%%%%%%%%%%%%%%%%%%
Earlier studies of the dipole model showed that a wide variety of DIS
data can be described with only a few assumptions.  In particular, it
was demonstrated that inclusive DIS can be described together with
inclusive charm production and exclusive diffractive vector meson
photo- and electro-production.  Especially the bSat dipole model is
very successful in describing the exclusive production of J/$\psi$,
$\phi$, $\rho$, and photon (DVCS) production at
HERA. %\cite{Kowalski:2006hc}.
Here we only give a short overview of the bSat model in $ep$ before we
discuss its extension to $e$A collisions.  For a detailed discussion
on the bSat model see \cite{Kowalski:2006hc}.

%%%%%%%%%%%%%%%%%%%%%%%%%%%%%%%%%%%%%%%%%%%%%%%%%%%%%%%%%%%%%%%%%%%%%%%%%
\subsection{A brief description of the bSat dipole model in diffractive $ep$}
\label{bSatEp}
\begin{figure}
    \begin{center}
        \includegraphics[width=\columnwidth]{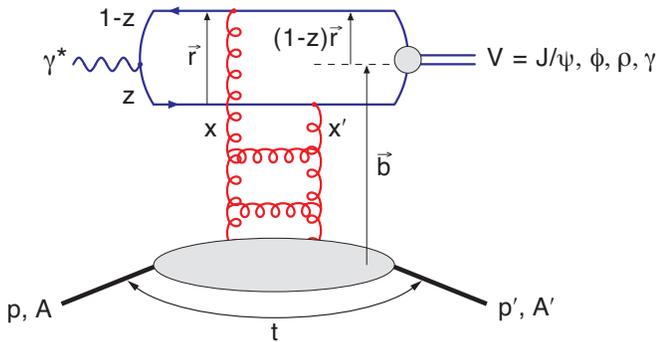}
    \end{center}
    \caption{\label{fig:dipole} (Color online) A schematic picture of the dipole
        model and its variables. See text for details.}
\end{figure}

%%%%%%%%%%%%%%%%%%%%%%%%%%%%%%%%%%%%%%%%%%%%%%%%%%%%%%%%%%%%%%%%%%%%%%%%
The amplitude for producing an exclusive vector meson or a real photon
diffractively in DIS can be written as:
\begin{eqnarray}
  \mathcal{A}_{T,L}^{\gamma^*p\rightarrow Vp}(x,Q,\Delta) =
  i\int{\rm d} r\int\frac{{\rm d} z}{4\pi}\int{\rm d}^2{\bf b}\left(\Psi^*_V\Psi\right)(r, z)
\nonumber &~&\\\times
2\pi rJ_0([1-z]r\Delta) 
  e^{-i{\bf b}\cdot{\Delta}}\frac{{\rm d}\sigma_{q\bar q}^{(p)}}{{\rm d}^2{\bf b}}(x, r, {\bf b})
  ~~&~&
    \label{eq:ttepamplitude}
\end{eqnarray}
where $T$ and $L$ represent the transverse and longitudinal
polarizations of the virtual photon, $r$ is the size of the dipole,
$z$ the energy fraction of the photon taken by the quark,
$\Delta=\sqrt{-t}$ is the transverse part of the four-momentum
difference of the outgoing and incoming proton, and ${\bf b}$ is the
impact parameter of the dipole relative to the proton (see
Fig.~\ref{fig:dipole}).  $(\Psi^*_V\Psi)$ denote the wave-function
overlap between the virtual photon and the produced vector meson.  In
this paper we use the "boosted Gaussian" wave-overlap with the
parameters given in \cite{Kowalski:2006hc}.

The dipole cross-section
${\rm d}\sigma_{q\bar q}^{(p)}/{\rm d}^2{\bf b}(x, r, {\bf b})$ is
defined as:
\begin{eqnarray}
    \frac{{\rm d}\sigma_{q\bar q}^{(p)}}{{\rm d}^2{\bf b}}
    (x, r, {\bf b})\equiv 2\mathcal{N}^{(p)}(x, r, {\bf b})=
    2[1-\Re(S)]
    \label{eq:dipolecs}
\end{eqnarray}
The first equality is the optical theorem, and we make the
approximation of only using the real part of the $S$-matrix for
the definition of the scattering amplitude $\mathcal{N}$,
which then becomes a real number between 0 and 1.
Here $(p)$ denotes proton.

In the bSat model the scattering amplitude is:	
\begin{eqnarray}
  \mathcal{N}^{(p)}(x, r, b)=
  1-e^{-\frac{\pi^2}{2N_C}r^2\alpha_S(\mu^2)xg(x,\mu^2)T(b)}
  \label{eq:bsatep}
\end{eqnarray}
where $\mu^2=4/r^2+\mu_0^2$ and $\mu_0^2$ is a cut-off scale in the
DGLAP evolution of the gluons.  The initial gluon density $xg(x,
\mu_0^2)=A_gx^{-\lambda_g}(1-x)^{5.6}$.  The nucleon profile function
$T(b)=1/(2\pi B_G)\exp(-b^2/(2B_G))$. All parameter values are
determined through fits to HERA data \cite{Kowalski:2006hc}.  For all
results in this paper, we use $B_G=4~{\rm GeV}^{-2}$, $\mu_0^2=1.17$
GeV$^2$, $\lambda_g=0.02$, and $A_g=2.55$. Also, the four lightest
quark masses are treated as parameters in the model, and are taken to
be: $m_u=m_d=m_s=0.14$ GeV, $m_c=1.4$ GeV.  It should be noted that
bSat is a model of multiple two-gluon exchanges at leading log, but
some nexto-to-leading order effects are taken into account by the running of the strong
coupling.

The total diffractive $\gamma^*p$ cross-section for this process is:
\begin{eqnarray}
  \frac{{\rm d}\sigma^{\gamma^*p}}{{\rm d}t}=\frac{1}{16\pi}
  \left|\mathcal{A}(x, Q^2, t)\right|^2
%  \nonumber
\end{eqnarray}

%%%%%%%%%%%%%%%%%%%%%%%%%%%%%%%%%%%%%%%%%%%%%%%%%%%%%%%%%%%%%%%%%%%%%%%%
\subsection{Extending the bSat model from $ep$ to $e$A}
\label{extendToEa}
%%%%%%%%%%%%%%%%%%%%%%%%%%%%%%%%%%%%%%%%%%%%%%%%%%%%%%%%%%%%%%%%%%%%%%%%

%%%%%%%%%%%%%%%%%%%%%%%%%%%%%%%%%%%%%%%%%%%%%%%%%%%%%%%%%%%%%%%%%%%%%%%%
The explicit impact parameter dependence of the bSat model makes it
especially well suited for the description of processes in $e$A
collisions. The $b$ dependence allows one to model the nucleus as a
collection of nucleons according to a given nuclear transverse density
distribution, e.g.~the Woods-Saxon function.  To this end we make two
observations.  Firstly, at small $x$, the life-time of the dipole
% , $eqn$,
is so large that the dipole traverses the full longitudinal extent of
the nucleus.  As a consequence the nucleus can effectively be treated
as a two-dimensional object in the transverse plane.  Also, when the
gluon's momentum fraction of the hadron is small, its wavelength in
the light-cone direction $x^-$ becomes so large, that it coherently
probes the whole nucleus at $x\ll A^{-1/3}/(M_NR_p)\sim 10^{-2}$,
where $M_N$ is the mass of the nucleus and $R_p$ is the proton radius.
Consequently, the information about which nucleon the gluon belongs to
is lost, and the exact position of each nucleon within the nucleus is
not an observable.  In order to calculate the cross-section correctly
the average over all possible states of nucleon configurations has to
be taken:
\begin{eqnarray}
    \frac{{\rm d}\sigma_{\rm total}}{{\rm d} t}=
    \frac{1}{16\pi}\left<\left|\mathcal{A}(x, Q^2, t, \Omega)\right|^2\right>_\Omega 
    %\nonumber
\end{eqnarray}
where $\Omega$ denotes nucleon configurations.

One defines two different kinds of diffractive events in $e$A:
coherent and incoherent. 
In the Good-Walker picture
\cite{Good:1960ba} the
incoherent cross-section is proportional to the variance of the
amplitude with respect to the initial nucleon configurations $\Omega$
of the nucleus:
\begin{eqnarray}
    \frac{{\rm d}\sigma_{\rm incoherent}}{{\rm d} t}&=&
    \frac{1}{16\pi}\bigg(\left<\left|\mathcal{A}(x, Q^2, t, \Omega)\right|^2\right>_\Omega
\nonumber\\&~&-
\left|\left<\mathcal{A}(x, Q^2, t, \Omega)\right>_\Omega\right|^2\bigg) 
%  \nonumber
\end{eqnarray}
where the first term on the R.H.S is the total diffractive cross-section and the
second term is the coherent part of the cross-section.

When extending the bSat model from $ep$ to $e$A we will use the independent
scattering approximation to construct the scattering amplitude for nuclei:
\begin{eqnarray}
    1-\mathcal{N}^{(A)}(x, {\bf r}, {\bf b})=\prod_{i=1}^A
    \left(1-\mathcal{N}^{(p)}(x, {\bf r}, |{\bf b}-{\bf b}_i|)\right)
    \label{eq:eptoea}
\end{eqnarray}
where ${\bf b}_i$ is the position of each nucleon in the nucleus in the 
transverse plane. 
We assume that the positions of the nucleons are distributed according
to the 3-dimensional Woods-Saxon function projected onto the transverse
plane. 
For details see Appendix \ref{app:Woods-Saxon}. 

Combining equations \eqref{eq:dipolecs}, \eqref{eq:bsatep} and
\eqref{eq:eptoea} the bSat scattering amplitude for $e$A becomes:
\begin{eqnarray}
	\frac{1}{2}	
  \frac{{\rm d}\sigma_{q\bar q}^{(A)}}{{\rm d}^2{\bf b}}(x, r, {\bf b}, \Omega)=&~&
  %2\bigg[
    \\
  1-\exp\bigg(-\frac{\pi^2}{2N_C}&r^2&    
    \alpha_S(\mu^2)xg(x,\mu^2) 
    \sum_{i=1}^AT(|{\bf b}-{\bf b}_i|)\bigg)
    %\bigg]
    .
  \label{eq:bSateA}
  \nonumber
\end{eqnarray}
Note that the dependence on nucleon configurations $\Omega$
in the amplitude is entirely contained in this dipole cross-section.

%%%%%%%%%%%%%%%%%%%%%%%%%%%%%%%%%%%%%%%%%%%%%%%%%%%%%%%%%%%%%%%%%%%%%%%%
\subsubsection{The incoherent, coherent, and total diffractive cross-sections}
%%%%%%%%%%%%%%%%%%%%%%%%%%%%%%%%%%%%%%%%%%%%%%%%%%%%%%%%%%%%%%%%%%%%%%%%
In order to obtain the total diffractive cross-section and its coherent part, the second and
first moments of the amplitude have to be calculated respectively. For the first
moment there is a closed expression for the average of the
dipole cross-section \cite{Kowalski:2003hm}:
\begin{eqnarray}
  \left<\frac{{\rm d}\sigma_{q\bar q}}{{\rm d}^2{\bf b}}\right>_\Omega = 
  2\left[1-\left(1-\frac{T_A({\bf b})}{2}\sigma^p_{q\bar q}\right)^A\right]
  \label{eq:analytical}
\end{eqnarray}
where $\sigma^p_{q\bar q}$ is the $ep$ dipole cross-section, eq.~(\ref{eq:bsatep}),
integrated over the impact parameter, and $T_A$ is the profile of the Woods-Saxon potential
in transverse space. 

\begin{figure*}
    \includegraphics[width=0.7\paperwidth]{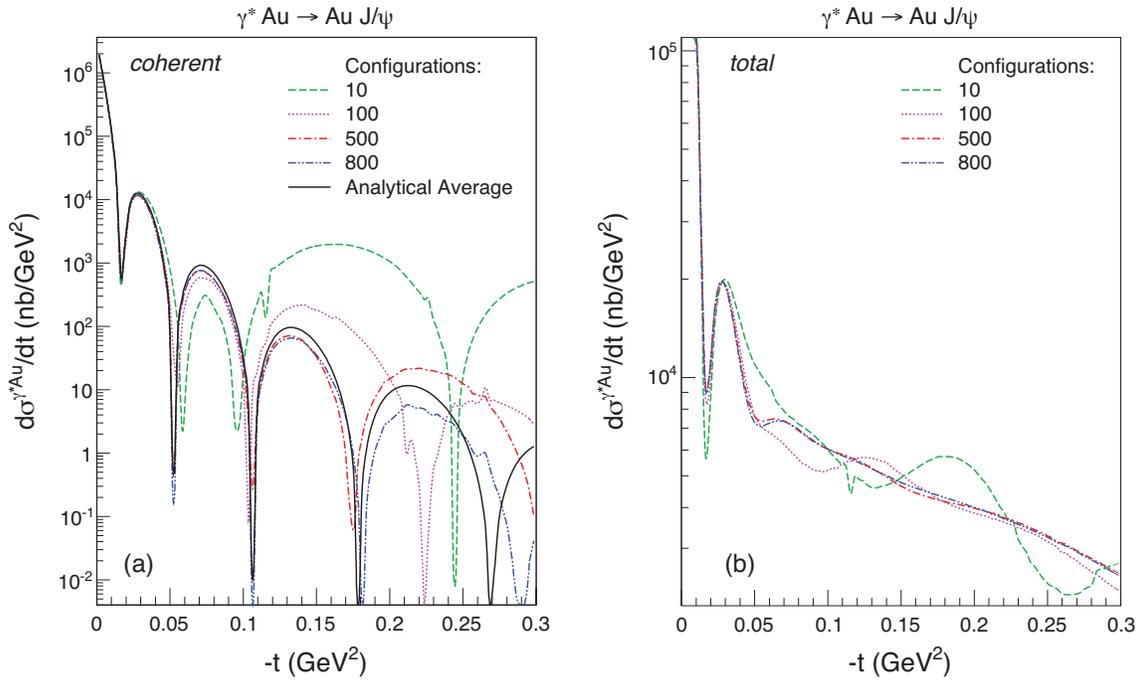}
    \caption{\label{fig:convergence} (Color online) (a) The resulting coherent and (b)
        total cross-section for
        $\gamma^*A\rightarrow\gamma^*J/\psi A$, averaged over 10, 100,
        500 and 800 configurations.  As reference, the coherent
        analytical average described by eq.~\eqref{eq:analytical} is
        also shown.}
\end{figure*}

For the second moment of the amplitude, no analytical expression
exists.  Similarly as in \cite{Kopeliovich:2001xj}, 
we derive it by defining an average of an observable 
$\mathcal{O}(\Omega)$ over nucleon configurations $\Omega_i$ by:
\begin{eqnarray}
    \left<\mathcal{O}\right>_\Omega=\frac{1}{C_{\rm max}}\sum_{i=1}^{C_{\rm max}}\mathcal{O}(\Omega_i).
    \label{eq:average}
\end{eqnarray}
For a large enough number of configurations $C_{\rm max}$ the sum on
the R.H.S.  will converge to the true average. For the total
diffractive cross-section one gets:
\begin{eqnarray}
    \frac{{\rm d}\sigma^{\gamma^*{\rm A}}}{{\rm d}t}(x, Q^2, t) = 
    \frac{1}{16\pi}\frac{1}{C_{\rm max}}\sum_{i=1}^{C_{\rm max}}\left|\mathcal{A}(x, Q^2, t, \Omega_i)\right|^2.~~~~
%    \nonumber
\end{eqnarray}
For large $t$ the variance is several orders of magnitude larger than
the average.  This means that the convergence of the sum in
eq.~\eqref{eq:average} becomes extremely slow, as demonstrated in
Fig.~\ref{fig:convergence}(a), where we show the coherent
cross-section resulting from averaging over 10, 100, 500, and 800
configurations. As a comparison the "analytical average", i.e.
eq.~\eqref{eq:analytical} is also shown. As can be seen, not even 800
configurations are enough for convergence at $-t>0.15$.

The convergence of the second moment of the amplitude is shown in
Fig.~\ref{fig:convergence}(b).  We conclude that around 500
configurations are needed to obtain a good description of the
cross-section for $-t<0.3$.

%%%%%%%%%%%%%%%%%%%%%%%%%%%%%%%%%%%%%%%%%%%%%%%%%%%%%%%%%%%%%%%%%%%%%%%%
\subsubsection{A non-saturated bSat model.}
\label{bNonSat}
%%%%%%%%%%%%%%%%%%%%%%%%%%%%%%%%%%%%%%%%%%%%%%%%%%%%%%%%%%%%%%%%%%%%%%%%
Saturation is introduced in the bSat model through the exponential
term in the scattering amplitude (eq.~\eqref{eq:bsatep}).  In order to
study the effects of saturation on the production cross-section we
construct a non-saturated version of the bSat model by linearizing
the dipole cross-section.  It should be noted that there is no taming
of the rise of the cross-section for small $x_{I\!\!P}$ or large
dipole radii in this case, and studies are only valid where
$\beta=x_{I\!\!P}/x_{\rm Bj}$ is large. For exclusive diffraction this
is equivalent to keeping $Q^2$ large. Any other way to impose a limit
on the rise of the cross-section, e.g.~through a cut-off, inevitably
also imposes some form of saturation into the formalism.

In the proton case, the bNonSat dipole cross-section is obtained by
keeping the first term in the expansion of the exponent in the bSat
dipole cross-section \cite{Kowalski:2003hm}:
\begin{eqnarray}
  \frac{{\rm d}\sigma_{q\bar q}^{(p)}}{{\rm d}^2b}
  =\frac{\pi^2}{N_C}r^2\alpha_s(\mu^2)xg(x, \mu^2)T(b).
%  \nonumber
\end{eqnarray}

In the case of a nucleus the dipole cross-section becomes:
\begin{eqnarray}
  \frac{{\rm d}\sigma_{q\bar q}^{(A)}}{{\rm d}^2b}
  =\frac{\pi^2}{N_C}r^2\alpha_s(\mu^2)xg(x, \mu^2)\sum_{i=1}^AT(|{\bf b}-{\bf b}_i|)
  \label{eq:bNonSateA}
\end{eqnarray}
and the coherent part of the bNonSat cross-section can be obtained through
the average:
\begin{eqnarray}
  \left<\frac{{\rm d}\sigma_{q\bar q}^{(A)}}{{\rm d}^2b}\right>_\Omega =
  \frac{\pi^2}{N_C}r^2\alpha_s(\mu^2)xg(x,\mu^2)AT_A(b).
  \label{eq:nosatCoherent}
\end{eqnarray}
The parameters we use for the bNonSat model were obtained in \cite{Kowalski:2003hm},
by fits to HERA data.
They are: $B_G=4$ GeV$^{-2}$, $\mu_0^2=0.8$ GeV$^2$, $\lambda_g=-0.13$, 
and $A_g=3.5$. The bNonSat quark masses are:
$m_u=m_d=m_s=0.15$ GeV, $m_c=1.4$ GeV.

\begin{figure}
  \includegraphics[width=\columnwidth]{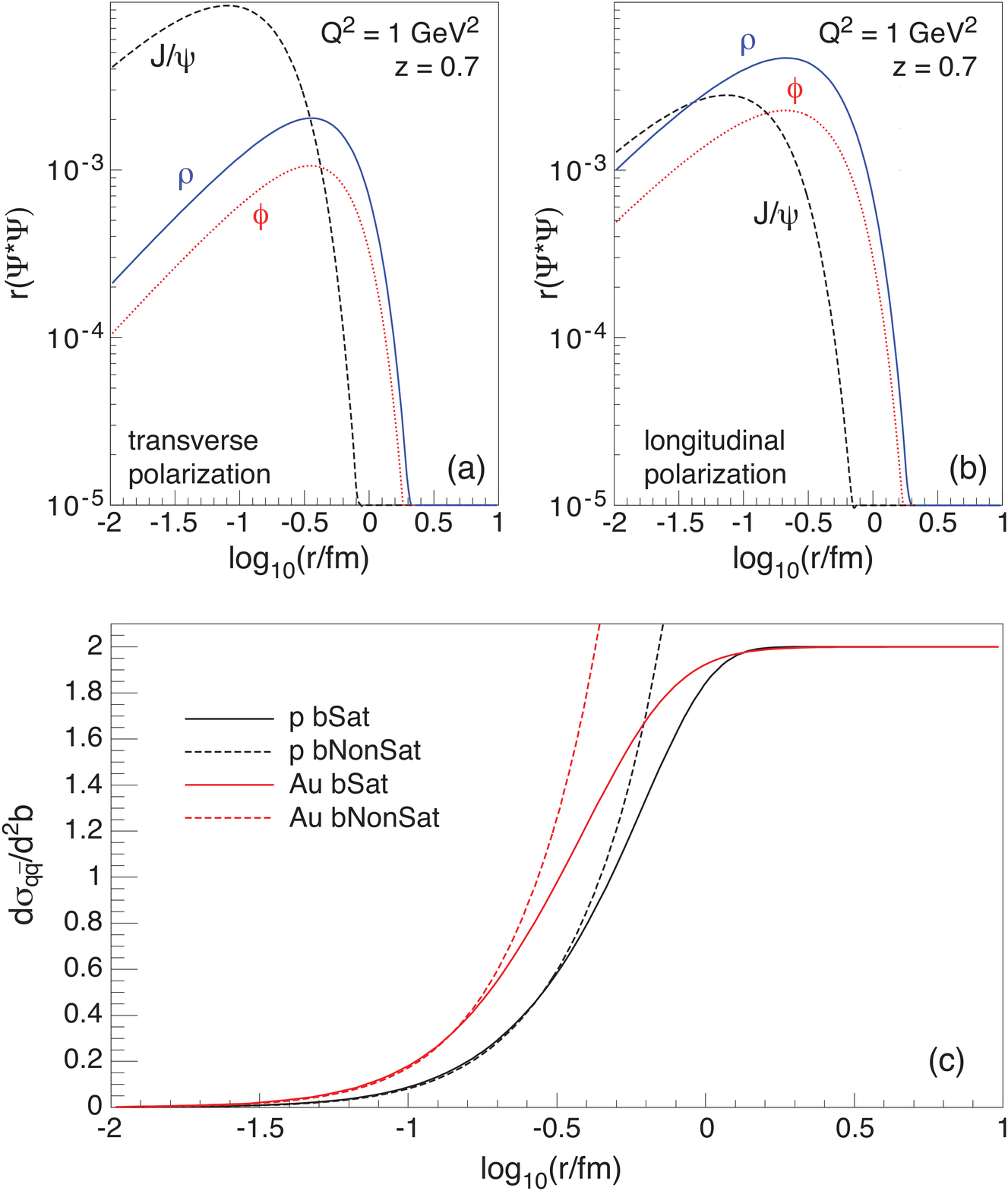}
  \caption{\label{fig:dipoleCrossSectionA} (Color online)
    In (a) and (b) the wave-overlap
    between the virtual photon and produced vector mesons are shown for transverse
    and longitudinal polarizations respectively, as functions of dipole radius $r$. 
    In the third panel the dipole cross-section is shown as a function of $r$,
    with bSat (solid) and bNonSat (dashed) for protons (black) and gold ions (red/grey).}
\end{figure}

Figures \ref{fig:dipoleCrossSectionA} (a) and (b) shows the
wave-overlap $(\Psi^*_V\Psi)$ between the virtual photon and produced
vector mesons as a function of dipole size $r$, for transverse and
longitudinal polarizations of the photon respectively. The
wave-overlap is taken at $Q^2=1$ GeV$^2$ and at $z=0.7$.  In
Fig.~\ref{fig:dipoleCrossSectionA} (c) we show the dipole
cross-section as a function of dipole size $r$.  In bSat the rise of
the cross-section at large $r$ is tamed in the model, while in bNonSat
it is allowed to rise uncontrollably.  Notice that despite the
uncontrolled rise of the dipole cross-section, the resulting
cross-section stays finite because of the steep fall of the
wave-overlap function at large $r$.  As can be seen in the figure, the
lighter (larger) vector mesons $\rho$ and $\phi$ are more sensitive to
saturation effects than heavier vector meson such as $J/\psi$.  For
$J/\psi$ the wave-overlap falls off so quickly at large $r$ that it is
an unsuitable probe for accessing the saturated regime, even for large nuclei.

%%%%%%%%%%%%%%%%%%%%%%%%%%%%%%%%%%%%%%%%%%%%%%%%%%%%%%%%%%%%%%%%%%%%%%%%
\subsubsection{Phenomenological corrections to the dipole cross-section}
%%%%%%%%%%%%%%%%%%%%%%%%%%%%%%%%%%%%%%%%%%%%%%%%%%%%%%%%%%%%%%%%%%%%%%%%
\label{corrections}
In the derivation of the dipole amplitude only the real part of the
$S$-matrix is taken into account.  The imaginary part of the
scattering amplitude can be included by multiplying the cross-section
by a factor $(1+\beta^2)$, where $\beta$ is the ratio of the imaginary
and real parts of the scattering amplitude.  It is calculated using
\cite{Kowalski:2006hc}:
\begin{eqnarray}
    \beta=\tan\left(\lambda\frac{\pi}{2}\right),~{\rm where}~
    \lambda\equiv\frac{\partial\ln\left(\mathcal{A}_{T,L}^{\gamma*p\rightarrow Vp}
    %(x,Q,\Delta)
    \right)}
                      {\partial\ln(1/x)}.
%  \nonumber
\end{eqnarray}

In the derivation of the dipole amplitude, the gluons in the two-gluon
exchange in the interaction are assumed to carry the same momentum
fraction of the proton or nucleus. To take into account that they
carry different momentum fractions, a so-called skewedness correction
is applied to the cross-section by multiplying it by a factor
$R_g(\lambda)$ defined by \cite{Kowalski:2006hc}:
\begin{eqnarray}
    R_g(\lambda)=\frac{2^{2\lambda+3}}{\sqrt{\pi}}\frac{\Gamma(\lambda+5/2)}{\Gamma(\lambda+4)}
%  \nonumber
\end{eqnarray}
where $\lambda$ is defined as above. Note that this definition of
skewedness-correction for the bSat model is slightly different from
the one used in \cite{Kowalski:2006hc}, but follows the description in
\cite{Lappi:2010dd}.

These corrections are important for describing HERA data, 
where the models are valid the corrections are typically around 60\% of the cross-section, out of which
the skewedness correction amounts to around 45\%.
The corrections grow dramatically in the large $x$ range outside the validity of the models,
where $x>10^{-2}$.
%%%%%%%%%%%%%%%%%%%%%%%%%%%%%%%%%%%%%%%%%%%%%%%%%%%%%%%%%%%%%%%%%%%%%%%%
\subsection{Computing the $e$A cross-sections}
%%%%%%%%%%%%%%%%%%%%%%%%%%%%%%%%%%%%%%%%%%%%%%%%%%%%%%%%%%%%%%%%%%%%%%%%
The differential $ep$ and $e$A cross-sections for exclusive
diffractive processes cannot be calculated analytically.  In order to
obtain numerical solutions we have written a computer program to
sample and average over nuclear configurations.  This program is also
the core of a novel event generator, Sar{\it t}re, which is briefly
described in Appendix \ref{Sartre}.

The total differential cross-section is:
\begin{eqnarray}
  \frac{{\rm d}^3\sigma_{\rm total}}{{\rm d}Q^2{\rm d}W^2{\rm d}t}=
   \sum_{T, L}\frac{R_g^2(1+\beta^2)}{16\pi}
  \frac{{\rm d}n_{T, L}^\gamma}{{\rm d}Q^2{\rm d}W^2}
  \left<|\mathcal{A}_{T, L}|^2\right>_\Omega~~ 
  \label{eq:totalCS}
\end{eqnarray}
where ${\rm d}n_{T, L}^\gamma/{\rm d}Q^2{\rm d}W^2$ is the flux
of transversely and longitudinally polarized virtual photons,
and the average over configurations $\Omega$ is defined
in eq.~\eqref{eq:average}.

The coherent part of the cross-section is:
\begin{eqnarray}
  \frac{{\rm d}^3\sigma_{\rm coherent}}{{\rm d}Q^2{\rm d}W^2{\rm d}t}=
  \sum_{T, L} \frac{R_g^2(1+\beta^2)}{16\pi}
  \frac{{\rm d}n_{T, L}^\gamma}{{\rm d}Q^2{\rm d}W^2}
  \left|\left<\mathcal{A}_{T, L}\right>_\Omega\right|^2~~
  \label{eq:coherentCS}
\end{eqnarray}
while the incoherent part is the difference between the total
and coherent cross-sections.

For the the second moment of the amplitude,
for each nucleon configuration $\Omega_i$, one need to calculate the integral:
\begin{eqnarray}
  \mathcal{A}_{T, L}(Q^2, \Delta, \xpom, \Omega_i)=
  \!\int r\dint r\frac{\dint z}{2}\dint^2{\bf b}
  \left(\Psi^*_V\Psi\right)_{T,L}(Q^2, r, z)
  &~& \nonumber \\ \times
  J_0([1-z]r\Delta)
  e^{-i{\bf b}\cdot{\bf \Delta}}
  \frac{\dint\sigma_{q\bar q}}{\dint^2{\bf b}}(\xpom,r,{\bf b}, \Omega_i)
  ~~~~~&~&
  \label{eq:moment2}
\end{eqnarray}
where the dipole cross-section is defined in eq.~\eqref{eq:bSateA} for
bSat and in eq.~\eqref{eq:bNonSateA} for bNonSat.  For $e$A, there is
no angular symmetry in ${\bf b}$ which makes this integral complex. We
average over 500 nucleon configurations, giving 1000 such integrals
for each point in phase-space.

For the first moment of the amplitude, the integral to calculate is:
\begin{eqnarray}
  \left<\mathcal{A}_{T,L}(Q^2, \Delta, \xpom)\right>_\Omega\!=\!
  \int\pi r\dint r\dint zb\dint b
  \left(\Psi^*_V\Psi\right)_{T, L}(Q^2, r, z)&~& \nonumber \\ \times
  J_0([1-z]r\Delta)
  J_0(b\Delta)
  \left<\frac{{\rm d}\sigma_{q\bar q}}{{\rm d}^2{\bf b}}\right>_\Omega(\xpom, r, b)~~~~&~&
  \label{eq:moment1}
\end{eqnarray}
where the average in the last term is defined in
eq.~\eqref{eq:analytical} for bSat and in eq.~\eqref{eq:nosatCoherent}
for bNonSat.

The dipole models described here are only valid for small values of
$x<10^{-2}$ and not too small values of $\beta\equiv x/\xpom$.  If
$\beta$ becomes too small the $q\bar q$ dipole becomes unphysically
large \cite{Kowalski:2008sa}. To rectify this one would need to
include higher Fock state dipoles, such as $q\bar q g$.

One should also note that the used dipole cross-section in Eq.~\eqref{eq:dipolecs}
and \eqref{eq:bsatep}, when integrated over the impact-parameter, yields unphysical 
results for large dipole radii:
\begin{eqnarray}
	\sigma^{(p)}(x, r) &=& 2\int\dint^2{\bf b}\mathcal{N}^{(p)}(x, r, b)\nonumber\\&=&
		4\pi B_G^2(\ln(G)-{\rm Ei}(-G)+\gamma_{\rm Euler})
\end{eqnarray}
where $G=(\pi^2r^2\alpha_{\rm S}(\mu^2)xg(x, \mu^2))/(2N_C2\pi B_G)$. For large $r$ the $\ln(r)$ contribution becomes dominant.
However, as demonstrated in Fig.~\ref{fig:dipoleCrossSectionA}, this growth has no 
effect on the actual production cross-sections (eq.~\eqref{eq:totalCS} and \eqref{eq:coherentCS}) due to the implicit cut-off of the wave-overlap at already moderate radii.

To nevertheless protect against this unphysical behavior, we introduce a cut-off in the dipole radius 
of $r<3$ fm for protons and $r<3R_0$ for nuclei, where $R_0$ is the nucleus' 
radius given in the Woods-Saxon parametrization. We varied the cut-off in a wide range and did not
observe any changes in the results presented here.

\section{Results}
\label{results}
%%%%%%%%%%%%%%%%%%%%%%%%%%%%%%%%%%%%%%%%%%%%%%%%%%%%%%%%%%%%%%%%%%%%%%%%
In order to verify that our numerical implementation reproduces
measured data, we repeated the comparison to the latest HERA data on
$\rho$, $\phi$, $J/\psi$, and DVCS. We find that both models, bSat and
bNonSat, describe HERA data well, within the experimental
uncertainties and within the kinematic validity of the models.  This
is not surprising since the $ep$ part is a repetition of previous work
(\cite{Kowalski:2003hm, Kowalski:2006hc}), although our
treatment of the skewness correction differs slightly.

%%%%%%%%%%%%%%%%%%%%%%%%%%%%%%%%%%%%%%%%%%%%%%%%%%%%%%%%%%%%%%%%%%%%%%%%
\subsection{Predictions for $e$A collisions}
To date, there exist no experimental data on diffractive vector meson
production in $e$A.  However, these measurements are integral parts of
the physics programs of future facilities such as the EIC
\cite{Deshpande:2012bu} and the LHeC \cite{Abramowicz:1998ii}.  We show results
for $J/\psi$ and $\phi$ production.  We let the $J/\psi$ mesons decay
into electron pairs, and the $\phi$ mesons into kaon pairs.  The
pseudo rapidity and momenta of these decay products are restricted to
$|\eta|<4$ and $p>1$ GeV, respectively. These cuts are made to limit
the predictions to an experimentally accessible region of phase-space.
We also limit the predictions to $x<10^{-2}$ and $Q^2>1$ GeV$^2$.  We
have simulated data corresponding to an integrated luminosity of $10$
fb$^{-1}$, with EIC beam energies of 20 GeV for the electron, and 100
GeV/$u$ for the ion beam.  This will amount to a few months of beam
operation.  The errors shown are statistical only.

\begin{figure*}
    \includegraphics[width=0.7\textwidth]{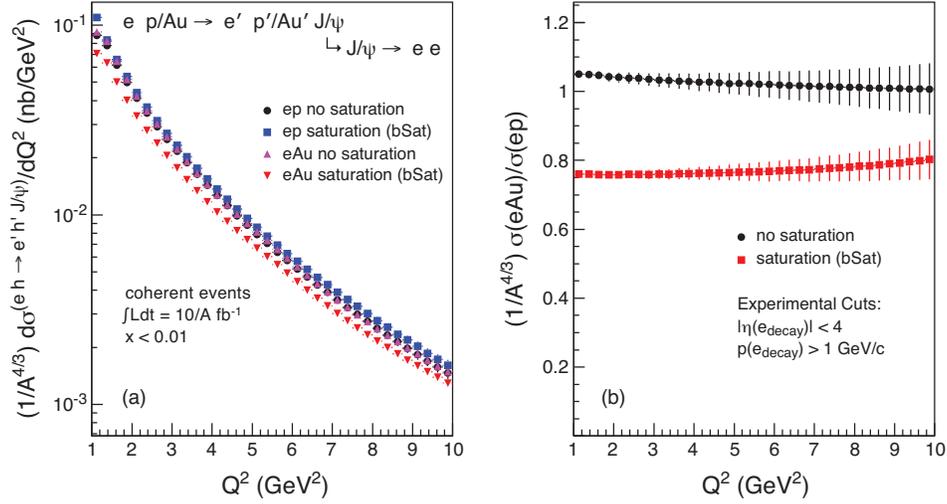}
    \caption{\label{fig:eic1}(Color online) (a) Cross-sections for $J/\psi$
        production differential in $Q^2$ for $ep$ and $e$Au collisions
        for both bSat and bNonSat dipole models. The cross-sections
        are scaled by 1/A$^{4/3}$. (b) Ratio of $e$A to $ep$
        cross-sections for both models.}
\end{figure*}

\begin{figure*}
    \includegraphics[width=0.7\textwidth]{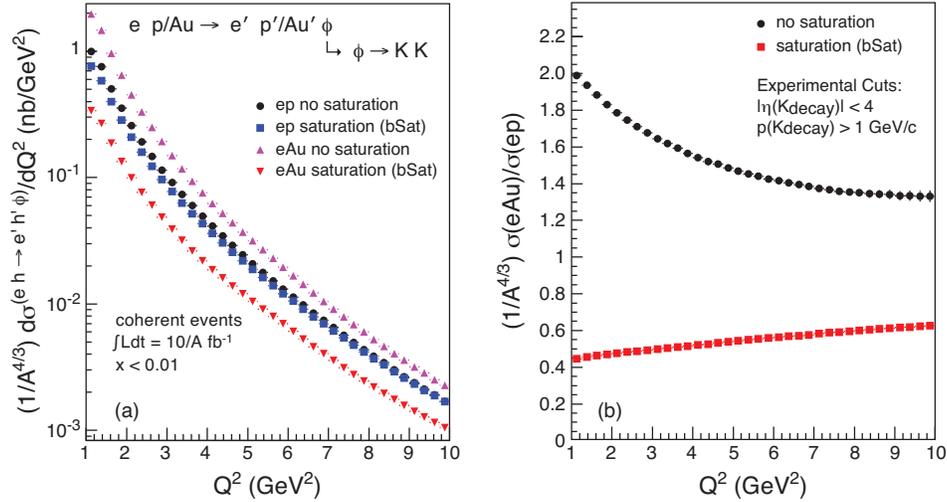}
    \caption{\label{fig:eic2}(Color online) (a) Cross-sections for $\phi$
        production differential in $Q^2$ for $ep$ and $e$Au collisions
        for both bSat and bNonSat dipole models. The cross-sections
        are scaled by 1/A$^{4/3}$. (b) Ratio of $e$A to $ep$
        cross-sections for both models.}
\end{figure*}

In Figs.~\ref{fig:eic1}(a) and
\ref{fig:eic2}(a) differential cross-sections with respect to $Q^2$ for
$J/\psi$ and $\phi$ production respectively are shown for both bSat
and bNonSat models.  The cross-sections are scaled by a factor
$A^{4/3}$. In the dilute limit (large $Q^2$) this scaling is expected
to hold for the integral of the coherent peak, which dominates the
cross-section, while deviations from it is due to the dense gluon
regime. 
In Figs.~\ref{fig:eic1}(b) and
\ref{fig:eic2}(b) the ratio of $ep$ to $eAu$
cross-sections are shown for both bSat and bNonSat.  As can be seen
there are significant differences between the two models, something
not observed at HERA. Also, the difference is larger for $\phi$
mesons. The reason for this is that the wave-function overlap between
the $\phi$ meson and virtual photon allows for larger dipoles than
that for $J/\psi$ (see Fig.~\ref{fig:dipoleCrossSectionA}).
Therefore, $\phi$ production can probe further into the dense gluon
regime and exhibits larger differences between bSat and bNonSat.


%%%%%%%%%%%%%%%%%%%%%%%%%%%%%%%%%%%%%%%%%%%%%%%%%%%%%%%%%%%%%%%%%%%%%%%%
\subsubsection*{Probing the spatial gluon distribution}
\begin{figure*}
    \includegraphics[width=0.7\paperwidth]{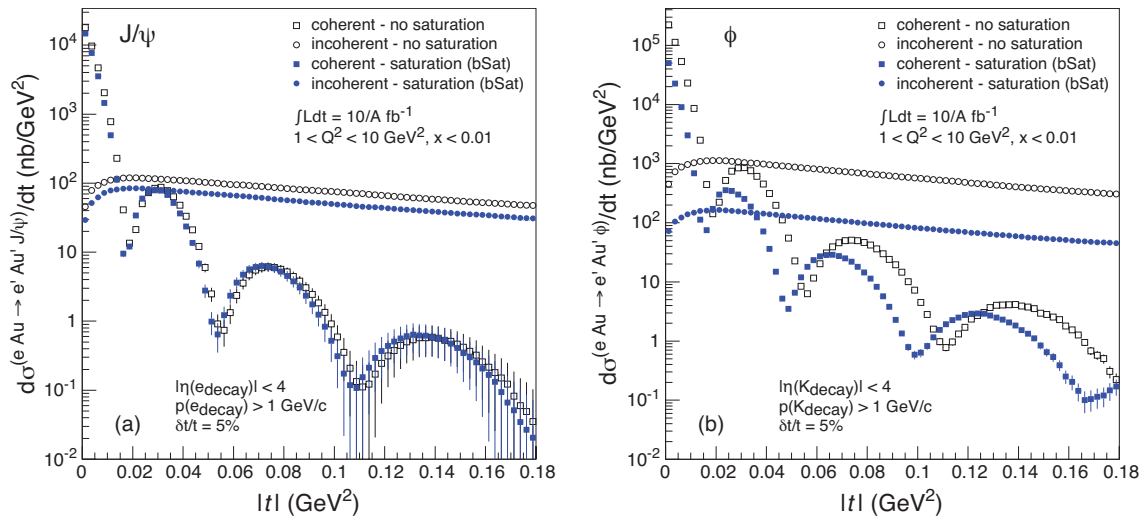}
    \caption{\label{fig:eic3} (Color online) Differential distributions with respect
        to $t$ for exclusive $J/\psi$ (a) and $\phi$ (b) for
        coherent and incoherent events.  Both bSat and bNonSat models
        are shown.}
\end{figure*}

In Fig.~\ref{fig:eic3} we show the differential cross-section with
respect to $t$, $\dint\sigma/\dint t$, for both $J/\psi$- and
$\phi$-meson production, again for both dipole models.  We assume a
conservative $t$-resolution of 5\%, which should be achievable by
future EIC detectors.  The statistical error bars shown correspond to
an integrated luminosity of 10 fb$^{-1}$.  As can be seen, the
coherent cross-section clearly exhibits the typical diffractive
pattern.  Also depicted in Fig.~\ref{fig:eic3} is the incoherent
cross-section, which is proportional to the lumpiness of the nucleus.
Experimentally the sum of the coherent and incoherent parts of the
cross-section is measured. Through the detection of emitted neutrons
(e.g.~by zero-degree calorimeters) from the nuclear break-up in the
incoherent case it should be experimentally feasible to disentangle
the two contributions unambiguously.

The coherent distributions in Fig.~\ref{fig:eic3} can be used to
obtain information about the gluon distribution in impact-parameter
space through a Fourier transform. In eq.~\eqref{eq:moment1}, the
first moment of the diffractive amplitude is a Fourier transform of
the dipole cross-section averaged over nucleon configurations, times
the wave-function overlap between the vector meson and virtual photon.
This represents a transformation from coordinate space to momentum
space ${\bf \Delta}$.  The coherent cross-section $\dint\sigma_{\rm
    coherent}/\dint t$ is proportional to the absolute square of this
amplitude.  Following \cite{Munier:2001nr}, we can regain the
impact-parameter dependence by performing a Fourier transform on the
amplitude.  The amplitude can be obtained by taking the square root of
the cross-section.  In order to maintain the oscillatory structure of
the amplitude we have to switch its sign in every second minimum.  We
call this modified amplitude $\sqrt{\dint\sigma_{\rm coherent}/\dint
    t}|_{\rm mod}$.  Its Fourier transform is:
\begin{eqnarray}
  F(b)=
  \frac{1}{2\pi}\int_0^\infty  
  \dint\Delta\Delta 
  J_0(\Delta b)
  \sqrt{\frac{\dint\sigma_{\rm coherent}}{\dint t}(\Delta)}\bigg|_{\rm mod},
  \label{eq:source}
\end{eqnarray}
which is a function of impact-parameter only.  In our models the
impact-parameter dependence comes from the transverse density function
$T_A(b)$.  For bNonSat, $F(b)$ is directly proportional to the input
density function, while for bSat the relation is more complex.

\begin{figure*}
  \begin{center}
    \includegraphics[width=0.75\paperwidth]{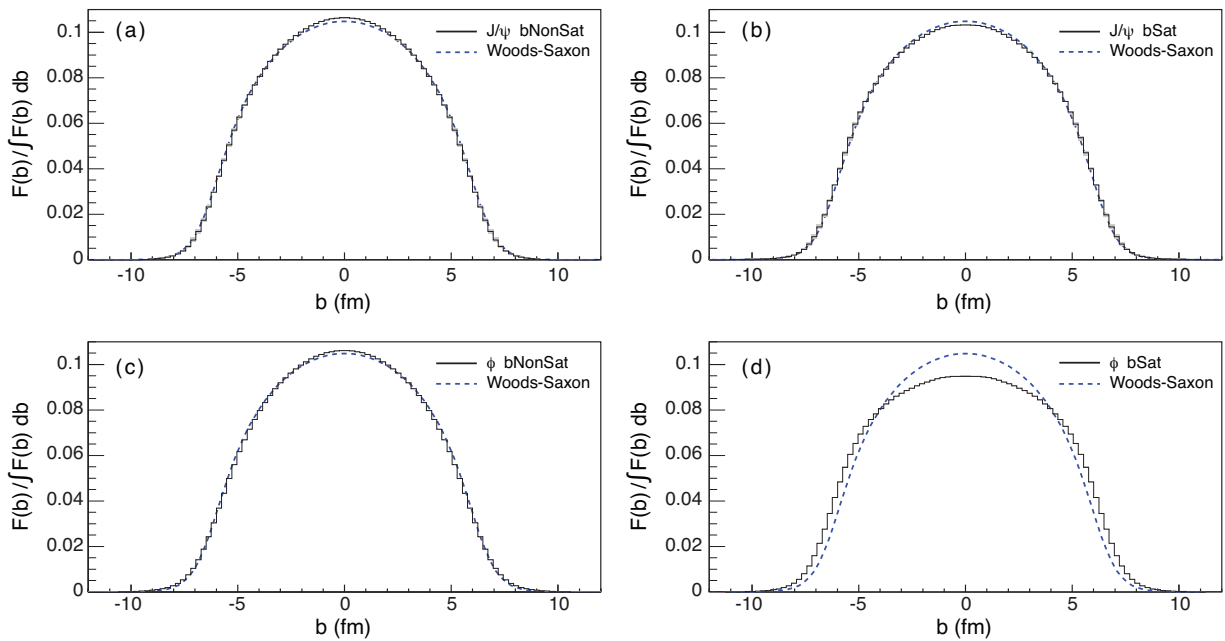}
  \end{center}
  \caption{\label{fig:source} (Color online) The Fourier transforms obtained from the 
  distributions in Fig.~\ref{fig:eic3} for $J/\psi$-mesons in (a) and (b)
  and $\phi$-mesons in (c) and (d).
  The results from both bSat and bNonSat are shown with error bands
  The input Woods-Saxon distribution is shown as a reference. 
  }
\end{figure*}
In Fig.~\ref{fig:source} we show the resulting Fourier transforms of
the coherent curves in Fig.~\ref{fig:eic3}, using the range where
$-t<0.36$ GeV$^2$.  The obtained distributions have been normalized to
unity.  For testing the robustness of the method, we used the
statistical errors in $\dint\sigma/\dint t$ to generate two enveloping
curves, $\dint\sigma/\dint t(t_i)\pm \delta(t_i)$, where $\delta$ is
the one sigma statistical error in each bin $t_i$. The curves are then
transformed individually, and the resulting difference defines the
uncertainty band on $F(b)$. Surprisingly, the uncertainties due to the
statistical error are negligible, and are barely visible in
Fig.~\ref{fig:source}.

As a reference we show (dotted line) the original input distribution
$T_A(b)$, which is the Woods-Saxon function integrated over the
longitudinal direction and normalized to unity.  The bNonSat curves
for $\phi$- and $J/\psi$-meson production reproduce the shape of the
input distribution perfectly as is expected since the bNonSat
amplitude is directly proportional to the input distribution.  For
bSat, the shape of the $J/\psi$ curve also reproduces the input
distribution, while the $\phi$ curve does not.  As explained earlier,
this is not surprising, as the size of the $J/\psi$ meson is much
smaller than that for $\phi$, which makes the latter more susceptible
to differences in the dipole cross-section between bNonSat and bSat,
as seen in Fig.~\ref{fig:dipoleCrossSectionA}.  We conclude that the
$J/\psi$ is better suited for probing the transverse structure of the
nucleus. 
However, by measuring $F(b)$ with both $J/\psi$- and $\phi$ mesons,
one can obtain valuable information on how sensitive the 
measurement is to non-linear effects. Thus, both measurements
are important and complementary to each other.
The results in Figure \ref{fig:source} provide a strong
indication that the EIC and the LHeC will be able to obtain the
nuclear spatial gluon distribution from the measured coherent
$t$-spectrum from exclusive $J/\psi$ production in $e$A, in a model
independent fashion.

\begin{figure}
  \begin{center}
    \includegraphics[width=0.9\columnwidth]{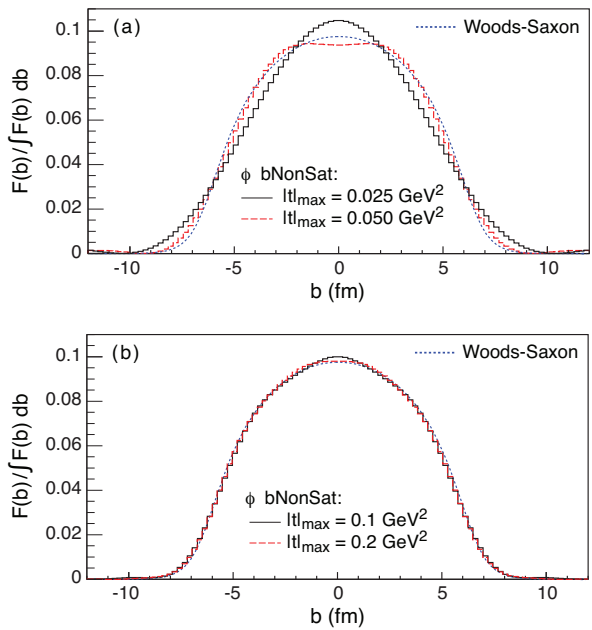}
  \end{center}
  \caption{\label{fig:sourceincr} (Color online) The Fourier transform of the $t$-spectrum
  of $\phi$-meson production in bNonSat, integrated to different
  upper values of $|t|$.}
\end{figure}
Strictly, the integral over $\Delta$ in eq.~\eqref{eq:source} should
be performed up to $\Delta=\infty$.  In Fig.~\ref{fig:sourceincr} we
demonstrate the effect of finite integration limits, using as an
example the $\phi$ meson curve.  We show the transformation for 4
upper values: $|t|_{\rm max}=\{0.025,~ 0.05,~0.1,~0.2\}$ GeV$^2$.  The
study shows a surprisingly fast convergence towards the input
Woods-Saxon distribution.

%%%%%%%%%%%%%%%%%%%%%%%%%%%%%%%%%%%%%%%%%%%%%%%%%%%%%%%%%%%%%%%%%%%%%%%%
\subsection{Ultra Peripheral Collisions}
%%%%%%%%%%%%%%%%%%%%%%%%%%%%%%%%%%%%%%%%%%%%%%%%%%%%%%%%%%%%%%%%%%%%%%%%
The calculations described in this paper can also be applied to Ultra
Peripheral Collisions (UPC) at hadron colliders, such as RHIC and the
LHC.  At very large impact-parameters between colliding hadrons the
long range electromagnetic force becomes dominant over short-range
QCD.  We substitute the electron's photon flux $\dint n^\gamma/\dint
Q^2\dint W^2$ in eq.~\eqref{eq:totalCS} with that from a proton or an
ion, as described in e.g.~\cite{Klein:1999gv}.

In Table \ref{tab:UPC} we list the predicted cross-sections for
$J/\psi$ mesons produced exclusively at RHIC energy in $p+p$, $p+$Au,
and Au+Au collisions.  Each cross-section is a sum of the two possible
photon directions in the events, such that symmetric beam particles
are multiplied by a factor 2, and the $p+$Au cross-section is the sum
of the photon coming from the proton and from the gold-ion
respectively. 
Especially for light mesons such as $\phi$, these studies might
provide new constraints for non-linear phenomena, such as saturation.
Measurements at existing hadron colliders are still limited in statistics
at the time of writing but more detailed measurements will become available soon.
The PHENIX experiment at RHIC measured the central UPC diffractive
$J/\psi$-production cross-section at $\sqrt{s}=200$ GeV, for $|\eta(J/\psi)|<0.35$ corresponding to $21<W<30$ GeV, when the $J/\psi$ decays into an electron pair \cite{Csanad:2009fz}.
The resulting cross-section is measured to be
$\dint\sigma/\dint y=76\pm33$(stat.)$\pm11$(syst.)${\rm \mu}$b. 

Our result is
$\dint\sigma/\dint y=118.5~{\rm \mu}$b, which is within the experimental uncertainty.
It should be noted that this measurement is at values of $\xpom\simeq 0.016$, which is bordering the validity range of the dipole model. In particular the phenomenological corrections to the diffractive cross-section described in section \ref{corrections} become large and are not under solid theoretical control.
\begin{table}
  \begin{center}
    \begin{tabular}{|c|c|}
      \hline
      Process & Cross-section (nb) \\
      \hline\hline
      $p+p$ & $0.716$ \\
      \hline
      $p+$Au & $0.666\cdot 10^3$ \\
      \hline
      Au+Au & $1.22\cdot 10^6$\\
      \hline
    \end{tabular}
  \end{center}
  \caption{\label{tab:UPC} Cross-sections of $J/\psi$ in UPC events at RHIC. 
    All cross-sections are for $\sqrt{s}=200~$GeV$/u$, 
    $10^{-6} \leq Q^2 \leq 1~$GeV$^2$, $4\leq W\leq 142~$GeV, $0\leq -t \leq 0.3~$GeV$^2$.}
\end{table}

%%%%%%%%%%%%%%%%%%%%%%%%%%%%%%%%%%%%%%%%%%%%%%%%%%%%%%%%%%%%%%%%%%%%%%%%
\section{Conclusions}
%%%%%%%%%%%%%%%%%%%%%%%%%%%%%%%%%%%%%%%%%%%%%%%%%%%%%%%%%%%%%%%%%%%%%%%%
We have presented a new method for calculating exclusive diffractive
vector meson and DVCS production in high energy $e$A collisions, based
on the dipole model.  This method is the first to describe incoherent
$e$A collisions without making approximations larger than those
already inherently present in the dipole model, for all values of $t$.  
In some parts of
phase-space, the cross-section is dominated by its incoherent part,
which is thus essential for making realistic predictions for future
$e$A experiments.  High energy $e$A collisions are expected to be
sensitive to non-linear saturation effects.  We have therefore
implemented our method in two dipole models: the bSat model and its
linearization the bNonSat model.

In Figs.~\ref{fig:eic1} and \ref{fig:eic2} we show that in an $e$A
collider, the two models are clearly distinguishable, which is not the
case in previous $ep$ experiments.  We also show that $\phi$-meson
production is considerably more sensitive to non-linear effects than
$J/\psi$-meson production.  This is due to the larger size of the
wave-function overlap for the $\phi$-meson.  In Figs.~\ref{fig:eic3}
and \ref{fig:source}, we show that one can probe the transverse
spatial gluon distribution of a nucleus by performing a Fourier
transform of the measured coherent $t$-spectrum. This method is very
robust with respect to statistical uncertainties and only requires
a range of $t\lesssim0.2$ GeV$^2$ for gold.  Due to its smaller wave
function, the $J/\psi$-meson is considerably more suitable for probing
the spatial gluon distribution than the lighter $\phi$-meson. In
Fig.~\ref{fig:eic3} we also show the incoherent $t$-spectrum, which is
directly proportional to the lumpiness of the initial nucleus. Our
method can also be used to calculate UPC events in present
hadron-hadron colliders. We describe central $J/\psi$ data from
the PHENIX experiment well within the experimental uncertainties.

%\begin{acknowledgments}
% Put your acknowledgments here.
%\end{acknowledgments}
\begin{acknowledgments}
The authors would like to thank Henri Kowalski, Tuomas Lappi, and Raju Venugopalan for their input and help, and the Open Science Grid consortium
for providing resources and support. 
This work was supported by the U.S. Department of Energy under Grant
No. DE-AC02-98CH10886.
%  Tom, Marko, Guillaume, Tuomas, Raju, Will, Henri, Francois, OSG etc.
\end{acknowledgments}
%%%%%%%%%%%%%%%%%%%%%%%%%%%%%%%%%%%%%%%%%%%%%%%%%%%%%%%%%%%%%%%%%%%%%%%%
\appendix
%%%%%%%%%%%%%%%%%%%%%%%%%%%%%%%%%%%%%%%%%%%%%%%%%%%%%%%%%%%%%%%%%%%%%%%%
\section{Generating a nucleon configuration according to the Woods-Saxon potential}
\label{app:Woods-Saxon}
%%%%%%%%%%%%%%%%%%%%%%%%%%%%%%%%%%%%%%%%%%%%%%%%%%%%%%%%%%%%%%%%%%%%%%%%
We generate the nucleus according to the Woods-Saxon distribution,
which is assumed to describe the number density of nucleons per volume
element, i.e.:
\begin{eqnarray}
    \frac{{\rm d}^3N}{{\rm d}^3{\bf r}}=
    \rho(r)=\frac{\rho_0}{1+e^\frac{r-R_0}{d}}
%    \nonumber
\end{eqnarray}
where $\rho_0$ is the central density, $R_0$ is the radius of the
nucleus and $d$ is the skin thickness which describes how fast the
potential falls off close to the edge of the nucleus.  The parameters
$\rho_0$, $R_0$ and $d$ have been measured for most nuclei in
low-energy experiments \cite{De Jager:1974dg}.

Our method for generating a nucleus is as follows:
\begin{enumerate}
\item We first generate the radial distribution of all nucleons in a
    given nucleus specimen according to:
    \begin{eqnarray}
        \frac{{\rm d}N}{{\rm d}r} = 4\pi r^2\rho(r),
%        \nonumber
    \end{eqnarray}
    and sort them in $r$.
\item We then generate the angular distributions uniformly in
    azimuthal angle, $\phi$, and polar angle, $\cos\theta$, one at a
    time beginning with the innermost nucleon.
\item If the newly generated nucleon position is within a core
    distance of $0.8$~fm from any other nucleon we regenerate $\phi$
    and $\cos\theta$, keeping the original $r$.  If this fails
    repeatedly, we drop the nucleus and restart from 1.
\item Finally, when all nucleons have been placed, the origin of the
    nucleus is shifted to its center of mass.
\end{enumerate}
%%%%%%%%%%%%%%%%%%%%%%%%%%%%%%%%%%%%%%%%%%%%%%%%%%%%%%%%%%%%%%%%%%%%%%%%
\section{Generating events with Sar{\it t}re}
\label{Sartre}
Sar{\it t}re is a novel Monte Carlo event generator, implementing the models
described in this paper. It generates exclusive events in diffractive
vector meson and DVCS production for $ep$ and $e$A collisions.

The master equation of Sar{\it t}re is eq.~\eqref{eq:totalCS}.  In the event
generator, this cross-section is simply used as a probability density
function from which a phase-space point in $Q^2$, $W^2$, and $t$ is
drawn. Given the beam energies and these three kinematic variables,
the final state of the event is fully defined except for the azimuthal
angle of the vector meson, which is uniformly distributed.

To determine the total cross-section in $e$A, the complex
four-dimensional integral described in eq.~\eqref{eq:moment2} has to
be calculated for each phase-space point 1000 times, which is
prohibitive for efficient event generation.  Therefore, we tabulate
the first and second moments of the amplitudes, for both
longitudinally and transversely polarized photons separately.  The
resulting look-up tables are three dimensional in $Q^2$, $W^2$ and
$t$.  There is a set of four look-up tables
($\left<|\mathcal{A}_T|^2\right>$, $|\left<\mathcal{A}_T\right>|$,
$\left<|\mathcal{A}_L|^2\right>$, $|\left<\mathcal{A}_L\right>|$) for
each species of produced vector meson or DVCS, and for each species of
nucleus.

When an event has been generated it is decided probabilistically
weather the event was coherent or incoherent by comparing the coherent
cross-section in eq.~\eqref{eq:coherentCS} with the total one.  In the
incoherent case we let the nucleus break up by assuming that the
diffractive mass $M_Y$ is distributed according to:
\begin{eqnarray}
  \frac{\dint N}{\dint M_Y^2}\propto\frac{1}{M_Y^2}.
%  \nonumber
\end{eqnarray}
Note that $M_Y$ cannot be uniquely determined from kinematics alone.
The corresponding excitation energy of the nucleus is:
\begin{eqnarray}
  E^*=(M_Y-m_n)\cdot A
%  \nonumber
\end{eqnarray}
We then use this excitation energy as input for \mbox{Gemini++}
\cite{Mancusi:2010tg}, a statistical model code which describes the nuclear
de-excitation, providing the break-up products from neutrons up to the
heaviest fragments.

% If in two-column mode, this environment will change to single-column format so that long equations can be displayed. 
%\begin{widetext}
%$$\mbox{put long equation here}$$
%\end{widetext}

% Figures should be put into the text as floats. 
% Use the graphics or graphicx packages (distributed with LaTeX2e).
% Put the label that you will use with \ref{} command in the braces of the \label{} command.
%
% \begin{figure}
% \end{figure}

% \begin{table}
% \end{tabular}
% \end{table}

%\bibliography{your-bib-file}

\end{document}